\documentclass[letterpaper,journal]{IEEEtran}

\usepackage[pdftex]{graphicx}
\usepackage{epstopdf}
\graphicspath{{./figures/}}
\usepackage{url}
\usepackage{cite}
\usepackage{color}
\usepackage{multirow}
\usepackage{algorithmic}

\usepackage{acronym}

\acrodef{GPS}{Global Positioning System}
\acrodef{OBD}{On-Board Diagnostics}
\acrodef{DoS}{Denial of Service}

\title{SenseMyCity: Crowdsourcing an Urban Sensor}

\author{%
Jo\~{a}o~G.~P.~Rodrigues\IEEEauthorrefmark{1},~\IEEEmembership{Student~Member,~IEEE,}
Ana~Aguiar\IEEEauthorrefmark{1},~\IEEEmembership{Member,~IEEE,}
Jo\~{a}o~Barros\IEEEauthorrefmark{1}\IEEEmembership{,~Senior~Member,~IEEE}%
\thanks{\IEEEauthorrefmark{1}Instituto de Telecomunica\c{c}\~{o}es, Departamento de Engenharia Eletrot\'{e}cnica e de Computadores, Faculdade de Engenharia da Universidade do Porto, Porto, Portugal}%
}

\begin{document}
\maketitle

\begin{abstract}

People treat smartphones as a second skin, having them around nearly 24/7 and constantly interacting with them.
Although smartphones are used mainly for personal communication, social networking and web browsing, they have many connectivity capabilities, and are at the same time equipped with a wide range of embedded sensors.
Additionally, bluetooth connectivity can be leveraged to collect data from external sensors, greatly extending the sensing capabilities.
However, massive data-gathering using smartphones still poses many architectural challenges, such as limited battery and processing power, and possibly connectivity costs. 

This article describes SenseMyCity\footnote{http://cloud.futurecities.up.pt/sensemycity/} (SMC), an Internet of Things mobile urban sensor that is extensible and fully configurable.
The platform consists of an app, a backoffice and a frontoffice.
The SMC app can collect data from embedded sensors, like GPS, wifi, accelerometer, magnetometer, etc, as well as from external bluetooth sensors, ranging from On-Board Diagnostics gathering data from vehicles, to wearable cardiac sensors.
Adding support for new internal or external sensors is straightforward due to the modular architecture.
Data transmission to our servers can occur either on-demand or in real-time, while keeping costs down by only using the configured type of Internet connectivity.
We discuss our experience implementing the platform and using it to make longitudinal studies with many users.
Further, we present results on bandwidth utilization and energy consumption for different sensors and sampling rates.
Finally, we show two use cases: mapping fuel consumption and user stress extracted from cardiac sensors.


\end{abstract}


\section{Introduction}
\label{sec:intro}




Crowdsourcing, crowdsensing or participatory sensing have recently become feasible (besides popular) ideas thanks to the rapidly growing number of affordable internet-enabled sensing devices---smartphones.
They are equipped with a wide range of embedded sensors, like GPS for location, magnetometer, accelerometer, gyroscope and most recently even barometer and temperature, as well as connectivity opportunities, like Bluetooth, WiFi, NFC and 3G.
Hence, they may become personal sensors in self-quantification applications that range from health to mobility~\cite{Boulos2011}, but also sensors for groups of people~\cite{Shilton2009} or for wide geographic areas~\cite{Murty2008a} through aggregation of the values sensed from multiple devices.
Obtaining large-scale dynamic datasets for a widespread geographic area, like collecting cartographic and environmental data (e.~g.\ street steepness, pavement type, average traffic delay, noise levels and pollution) in urban areas becomes a tractable problem if the sensing can be crowdsourced.



However, mobile phones have limited resources, and should nevertheless be available at all times.
Users will only collaborate and provide data if the offered incentives or services overcomes at least the possible monetary costs.
Data collection typically requires communication, from the user's device to a centralized system, which can have inherent transmission costs when using a payed data connection.
Furthermore, gathering data from a personal device such as a smartphone should take security and privacy issues into consideration.
The users should be in control of when their devices are gathering possibly sensitive data, who can access that data and for what reason.

%
%

In existing sensing platforms these aspects are typically tuned in a trade-off manner and hard-coded in the system according to their specific requirements, e.g.: reducing battery consumption by sacrificing sampling rate, or transmitting only when connected to a free data connection and disallowing real-time data transmissions.

In this article, we extend our previous work~\cite{Rodrigues2011} where we proposed an architecture for a data gathering system and a working prototype based on a netbook.
We now describe the design of a framework for a crowdsourced urban sensor that has achieved a maturity level that enables its large-scale deployment among common citizens.
The framework is comprised of a smartphone application that gather data from many internal and external sensors and transmit it securely to a backend server.
It is modular and fully configurable, being able to meet requirements on resources consumption, data usage and sampling rates, by simple configuration changes.
This effectively allows the trade-offs to be performed per usage scenario and even changed in real-time according to the environment, e.g. decreasing sampling rates if the battery goes below a configured level.
We use a two-tier servers architecture to store, process and visualize the massive amounts of data.



The main contributions of this paper are:
1) the requirement analysis of a smartphone-based crowdsourced urban sensor (Section~\ref{subsec:requirements}); 
2) the design of a sensing system that meets the requirements of both users and researchers (Section~\ref{sec:system});
3) a modular and configurable mobile Android OS application, that can be used in various sensing tasks (Section~\ref{sec:mobileapp});
4) a resource consumption study of the different sensors and sampling rates, in terms of energy and storage, providing us an estimation of the cost-of-knowledge of such sensing tasks (Section~\ref{sec:meetinginterests});
5) example projects that already use the system and gathered data (Section~\ref{sec:usecase}).



\subsection{User Requirements}
\label{subsec:requirements}

A framework for a data gathering system has to satisfy two main stakeholders: the participants that have to use and handle the gathering device; and the researchers or operators who wish to gather the data.
We designed our system with these users' interests in mind.

\textbf{Intuitive and low user-interaction required}: We aim for a massive data gathering system, so the gathering unit should require the least user interaction possible.

The gathering and sensing devices should be \textbf{non-intrusive} in order to maximize user utilization and avoid biased information.
If the user have a constant sensation that his environment is being monitored, he might have a different behavior and avoid using the system. 

\textbf{Lightweight and stable}: 
Mainly if the device is used for other tasks, the application should work seamlessly, without affecting the normal user operation.

The \textbf{Battery Consumption} is one of the most important factors.
The application should provide information about energy consumption of different configurations, and allow a user to configure the desired sensors.

Our application should support any type of internet connectivity.
However, to \textbf{avoid mobile communication costs}, the user should be in control of when communications should occur and which connection type can be used.

\textbf{Transparency}: The user interface should clearly state what the application is doing. 
No data should be gathered nor transmitted without the user's knowledge.

\textbf{Security and privacy} should also be tackled throughout the application.
Also, the user shall retain ownership over gathered data, with full access to his own data, including being able to download the full data-set or delete it.
Furthermore, the required user identification mechanism should not be directly reversible, nor accessible by researchers nor data analysis services, not even when processing clusters of data.


\subsection{Researcher Goals}

\textbf{Ubiquitous Availability}: The system should support widely used and economic platforms, avoiding the need for specific hardware requirements or expensive devices.

\textbf{Modular, flexible and extensible}: Allow a wide variety of sensors and be easily extended for new sensors and gathering devices.

\textbf{Reliable}: Preserve as much accuracy as possible, from the sensors to the storage facility, in order not to hinder any current or future use the data may have.


\textbf{Scalability}: 
To increase availability even under high demand, the system should be scalable and compatible with load balancing mechanisms.

\textbf{Storage Utilization}: The researchers should be provided an estimation of the bandwidth and storage required for a given sensing task.

\section{SenseMyCity Crowdsensor}
\label{sec:system}

To meet the above requirements, we separated our system design in four sections: system architecture including data structures, the communication protocol, a working implementation based on a Mobile Application and the required Back-End servers.

\subsection{System Architecture}
\label{subsec:multitier}

\begin{figure}[tb]
 \centering
 \includegraphics[width=0.7\linewidth]{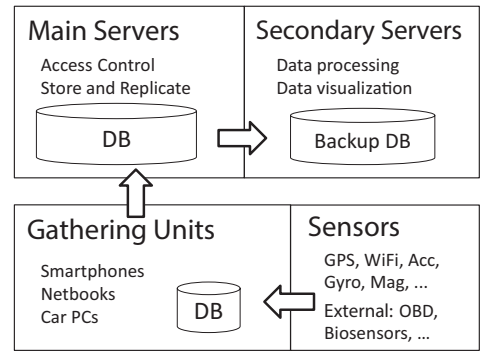}
 \caption{System Architecture}
 \label{figure:Arch}
\end{figure}

Our system is organized in a 3-tier architecture (Fig.~\ref{figure:Arch}), where each block has distinct requirements and functions.

The \textbf{gathering units} make the base block of the system, collecting data acquired by the sensors.
However modular, these units are responsible for synchronizing the data from the various sources they may have, to ensure quality and reliability of the data, and to encapsulate and transmit it to the main servers using our protocol.
The communication between these units and the main servers is performed using a machine-to-machine protocol based on JSON and standard encryption protocols, and is described in Section~\ref{sec:secure}.
We have implemented data collection programs in netbooks, car-PCs and have also created an application running on Android based smartphones that is further discussed in Section~\ref{sec:mobileapp}.

The \textbf{main servers} are the system core.
Their job is to receive the data from the units, store it, and replicate it asynchronously to secondary servers.
They should be reliable with a very high uptime, and offer large storage space.
To increase the uptime, the database storage engine should only perform write queries and simple indexed read queries that may be needed when creating or restoring a secondary server.
These main-servers can also be configured as a load-balancing or multi-master database if necessary due to increased write demand.
There are software solutions that provides multi-master architecture for relational databases supporting primary-keys collision-avoidance, and our system should be compatible with any of those solutions.


\textbf{Secondary servers} are responsible for provide any desired application level service, such as data processing and visualization, and to provide user defined privacy policies, anonymity and access control.
These services usually require higher bandwidth and complex database read queries that can leave any database engine unresponsive for some time.
The servers can either maintain a replica of the entire database or just a specialized subset of data, according to some project data requirements or to location of the data, such as a server dedicated to process the data from a specific city.

Our system should be compatible with any existing multi-server, replication or load-balancing database software, that are able to improve the uptime and scalability of the system.
However, we would like to emphasize the importance of a multi-server architecture (Fig.~\ref{figure:Arch}) in sensing systems, because the data gathering and storage, and the data analysis, processing and visualization, have completely distinct requirements: Storing the data takes little computational power, but should be as reliably and quick as possible; Analyzing or visualizing the data can be very computationally expensive, new experiments can block or destroy the whole database, but in most cases does not have tight deadlines.

This division in three blocks with two-tier servers also provides added security, since the only servers exposed to users or researchers are the secondary ones, which can be rolled-back at any time without loosing raw data.

\subsubsection{Data Structure}
\label{subsec:datastructure}

The used database structure is not fixed, being updated when new sensors or requirements are added to the system.
However, to increase the framework usability and compatibility between different projects and implementations, we defined a set of possible data types or tables.

\textbf{Session information}, with data about the recording session such as starting time, gathering unit identifiers, user and vehicle information, cryptography keys, and application version.
These tables should provide an identifier to be used by the rest of the database, such as user\_id, vehicle\_id and session\_id.
User identification exists for authentication and ownership reasons, but these are implemented using a one-way function such as hashing and these tables should not be accessible by any researcher or user.

\textbf{Gathered data} from sensors or other sources, such as questionnaires or device events, have distinct formats and thus can have their own data types and structure.
However, they should follow some structural rules to facilitate relational data-analysis.
To this end, every gathered data should be timestamped and identified by the current session\_id.
A common identifier makes session analysis and access-control much easier to implement.

These tables get one new row of data per second for typical periodic sensors.
Sensors with higher frequencies, such as accelerometer or gyroscope, are stored as an array.
A milliseconds column should be used by asynchronous sensors, such as an \ac{OBD} device where the delay between readings is not constant, or by sensors having a higher time precision, such as \ac{GPS}.
Questionnaires responses and device events, such as battery consumption status, can also be stored and sent to the server.

\textbf{Auxiliary data} are tables that contain non-personal data shared among many sessions or users.
Some examples are tables relating mac\_addresses with the WiFi essid, that save storage space and bandwidth by storing the repetitive and static information about the networks found.


\subsection{Reliable and Secure Data Gathering}
\label{sec:secure}

In this section we describe the communication protocol that handles the communication between gathering units and main servers.
It was designed to be efficient but secure, to protect the possibly sensitive data gathered from a users' devices.
The protocol guarantees confidentiality through encryption, integrity and reliability through compression and feedback, and an access control mechanism is responsible for authentication.

Our protocol encrypts every transmission using standard mechanisms, more precisely 4096-bits RSA public key cryptography to exchange per-session random 128-bits AES symmetric keys.


We implemented reliability through encrypted application layer feedback after successful storage of the data on the database.
Also, to minimize the resources utilization and damage done by lost packets, our server implementation works in a stateless mode, without keeping record of an internal communication state and treating every arriving packet the same way.
These features allow a gathering unit to transmit data in real-time or on-demand at arbitrary times, using any kind of network connection.
A unit can opt for the utilization of network friendly protocols (such as TCP) with a high-quality connection, while using other more efficient or packet oriented protocols (UDP) in other network environments, such as vehicular networks.

For authentication and access control we avoid account creation and password handling by using OpenID, an open and secure authentication method.
To access the data, a user has to login in the front-end server via OpenID, with the same e-mail configured and verified in the gathering unit.
To this end, the only user identification automatically stored on our servers is a normalized and hashed version of the user e-mail, which is not directly identifiable.
Furthermore, the table relating an user's hash to the internal user ID is protected, only accessible by the authentication mechanism and the DB Administrator.


The gathering unit is responsible for authenticating the user's OpenID when gathering or uploading data, but for efficiency we perform no user authentication on the server when storing the data.
This may allow a malicious user to replicate the data-gathering protocol or to fake the OpenID validation.
However, in this situation the attacker can only upload data under a fake user ID, and is not able to access any existing data from another user.
Other data forgery attacks are also impossible to prevent, e.g. almost every smartphone can be configured to return fake data from the integrated sensors.
A user can also alter gathered data just by changing the date and time of the mobile device's internal clock.

\begin{figure}[tb]
 \centering
 \includegraphics[width=0.7\linewidth]{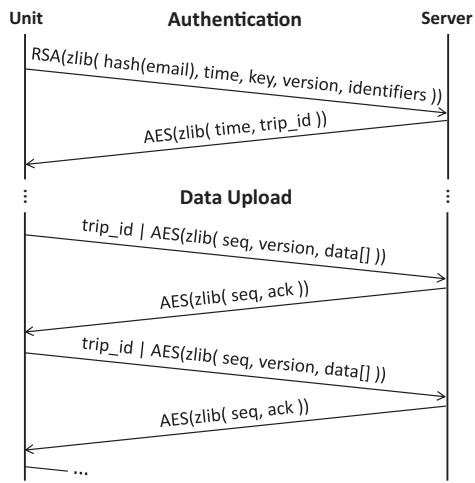}
 \caption{Protocol description, with session authentication and data transmission with feedback}
 \label{figure:protocol}
\end{figure}

We decided not to use available cryptographic protocols such as TLS or DTLS for simplicity reasons.
Even though those protocols are standard and with proven security, they offer many capabilities that are not required for applications like ours, such as Certificate or Cipher Specification Exchange.
By designing a custom protocol based on the same security standards, we were able to integrate our authentication phase in the first transmitted packet responsible for handshake and key exchange, reducing the number of overhead messages to zero (Fig.~\ref{figure:protocol}).
We also did not use any existing web-service standards or frameworks for the same reasons: complexity and overhead.
They typically bring architectural and design constraints like required Resource Identification, fixed message formats, or specific allowed communication protocol.
A data gathering platform should be able to deal with massive amounts of messages and data, making it very important to reduce required processing and bandwidth.
On the other hand, it only needs to support session authentication and data upload.

The protocol is divided in two parts (Fig.~\ref{figure:protocol}): a 2-way handshake and a data transmission phase with feedback.
The \textbf{2-way handshake} phase is responsible for session authentication and symmetric key exchange.
The gathering units should authenticate every sensing session by sending a packet.


\begin{itemize}
\item seq: a sequence number defined by the gathering unit.
Identifies the response packet, allowing the unit to send multiple authentication packets in parallel.

\item hash: this is the md5-hash of the user e-mail.

\item time: the unix timestamp of the start of the session.
This timestamp together with the user\_id make the session's primary key.

\item key: a per-session randomly-generated AES-128 symmetric key that should be used for the remainder of the session.

\item version: the application's version, to allow for future format upgrades with backwards compatibility.

\item identifiers: reserved for optional extra session information, such as identifiers of the gathering unit, sensors or vehicles.
\end{itemize}

Both the key and the identifiers of a session can be changed by the gathering unit at any time, by re-sending an authentication packet with the same hash (user ID) and time.
This allows the gathering unit to not store the aes\_key of each session and to update any identifier that may be unavailable at the start.

On receiving this packet, the server should initialize the session on the database and return the time and unique session identifier (session\_id) now reserved for that session's data.
Both fields are compressed and encrypted before transmission using the received symmetric key.

In the \textbf{data transmission phase}, the gathering unit creates a packet where the first 4 bytes correspond to the received session's unique ID, followed by the compressed and encrypted packet with the data to be transmitted.
The ID and the used encryption key must be the ones exchanged during the handshake, otherwise the server will discard the packet.

The sequence number is used to identify the feedback packets, allowing the unit to transmit the next data packets before receiving feedback from the previous one.
This asynchronous transmission can substantially increase the throughput in high latency connections.
The gathering unit is responsible for sequence number generations, tracking, and any necessary re-transmissions.

The feedback packet is only generated after the storage and flushing of the received data, and contains the number of rows received and successfully written to the DB.
This serves as guarantee of receipt and storage, making it safe to the gathering unit to delete the local data after receiving a positive feedback.

In our implementation we used JSON and zlib to serialize and compress the data before transmission, since they completely serves our purpose while being simple and lightweight. 
%

\subsection{Mobile Data Gathering Application}
\label{sec:mobileapp}


The Android application, labeled SenseMyCity, was developed to become a testbed for the FutureCities Project at Instituto de Telecomunica\c{c}\~oes.
The application is aimed at urban data gathering, taking advantage of the Android Smartphones’ embedded sensors.

In order to gain acceptance by a large audience in the general public, SenseMyCity was designed to run on a wide range of Android smartphones and require almost no configuration and user interaction.
The interface was designed to be intuitive and easy to use in any possible scenario.
A significant amount of time was invested on enhancing the stability and efficiency of the application to allow it to run properly across smartphones with different hardware resources.


\subsubsection{Interface}

The application is designed to have a simple and minimalistic interface, with just five big buttons available to start, pause or terminate the gathering session, to access the settings menu, and to synchronize the gathered data with the server.
A widget is also available to quickly start, pause or stop a session without leaving the smartphone's home screen.
A verbose interface is also available that shows the activated sensors and most recent gathered data.

The settings menu is very complete, since almost every possible parameter is configured from this interface.
However, the application was designed to be almost configuration free, with default settings that gather data from GPS, accelerometer and Wifi networks.
Settings include, among others, activation and deactivation of individual sensors, their desired sensing frequency, choosing between real-time data transfers or user-activated synchronization, and the allowed internet connection type to use on real-time data transmissions.
Furthermore, a researcher can fix or limit the configurations, like specifying active sensors and their sampling rates, prior do deployment.

When a sensing session is started, the application initializes the configured sensors and connects to the external devices.
The arriving data is buffered, stored and transmitted, as configured in the settings menu.
The application keeps collecting data in the background, without an active interface, but always showing a notification with its status.



\begin{figure}[tb]
 \centering
 \includegraphics[width=0.7\linewidth]{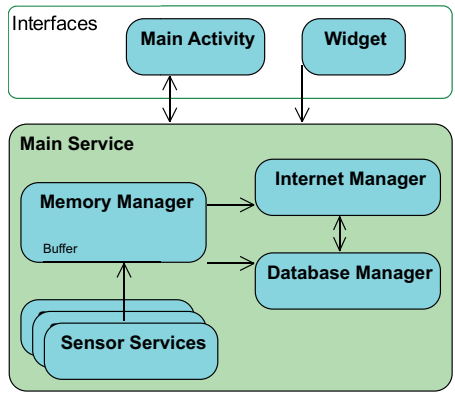}
 \caption{Application architecture depicting the various processing modules and information flow between them}
 \label{figure:apparch}
\end{figure}

\subsubsection{Architecture (Fig.~\ref{figure:apparch})}

The main service is responsible for controlling the whole application state and passing feedback to any active interface or notification.
This service starts gathering sessions by initializing the configured sensor threads and the required managers.
Every resource is executed in a separate thread, including the network connection, database access, each sensor, memory buffer and UI.
The main service also ensures that every sensor thread is correctly terminated and the data is flushed to local storage when the user terminates the session.

Each sensor thread communicates with the corresponding sensor, and receives, timestamps and sometimes converts the gathered data before sending it to the Memory Manager.
During tests, the main performance bottleneck of our mobile application proved to be the local SQLite storage, that can introduce a delay of a few seconds for a single database operation.
This manager buffers the data and allows the application to only require one SQLite database transaction every few seconds, while performing optional real-time synchronizations at the same time.

The Internet Manager implements the designed communication protocol for authenticating and uploading the data to our servers described in Section~\ref{sec:secure}.
All of the required cryptography and compression algorithms are available in the Android or java standard libraries.
The server RSA key is hardcoded on the application and thus can be changed anytime via an application update.
The symmetric keys are randomly generated when a session is started.

The application requires at least one google account to be configured in the android phone, leveraging in this way the built-in Google account verifier as the authentication method of the user e-mail, described in Section~\ref{sec:secure}.
This allows a user to login to our front-end using Google's OpenID authentication.

\subsubsection{Data Synchronization}


A common problem of data gathering systems is the synchronization and timing accuracy of the data.
To minimize the losses of time information, getting the device timestamp in milliseconds is the first performed task after every new sensor data.
This guarantees a minimum delay and jitter in time information.

We can further improve timing accuracy with offline processing, e.g., GPS data is stored containing both the satellites' and the smartphone's internal clock, and can thus be used to compare and correct the device's clock.




\subsection{Back-office Implementation}
\label{sec:implementation}


Our back-office implementation is running on two servers, implementing the proposed tiered architecture with one main server and one secondary server.

\begin{figure}[tb]
 \centering
 \includegraphics[width=0.7\linewidth]{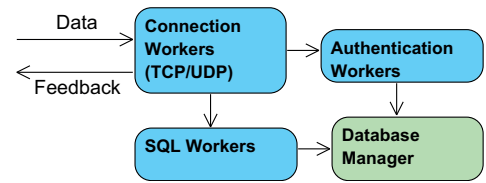}
 \caption{Back-end implementation}
 \label{figure:backarch}
\end{figure}

\subsubsection{Main Server (Fig.~\ref{figure:backarch})}
\label{subsec:backend}

On our sample implementation we use a lightweight java daemon to handle the requests and a SQL database for storage.
The daemon is composed by five threads: two Connection Workers listening for incoming packets, an Authentication Worker handling hand-shakes and key-management, an SQL Worker decrypting and decompressing received data packets, and a Database Manager performing any required database operation.
We have developed the server to use both PostgreSQL and MySQL local databases, but it is easily configurable to work with any other storage engine.

The java application listens on two distinct sockets for authentication or data packets, and in both UDP and TCP mode.
As stated in Section~\ref{sec:secure}, this application is stateless in both the authentication and data transmission phases, analyzing every received packet the same way.

The worker that handles authentication packets is responsible for decrypting them using the server private key, decompressing the data, and sending the SQL commands to the database manager.
This service discards and ignores any packet that cannot be decrypted using the server public key.
Every database request related to authentication is handled by the same DB connection to prevent lockups and race-conditions, running in an optimized worker thread.
This inserts or updates the corresponding session on the database, and returns the corresponding unique session\_id back to the gathering unit.
The authentication worker also caches recent session\_id\,-\,aes\_key combinations to avoid unnecessary database reads when receiving data packets.

Arriving data packets are handled by another worker thread, that starts by searching for the corresponding AES key, and discards the packet if no key is found or the decryption and decompression steps fail.
To maximize performance, this worker parses and separates the received data into optimized structures, before sending it to the database worker.
This database worker is a different one from the authentication phase, since it operates over different tables.
Again, feedback is only sent back to the unit when storage confirmation is received from the database.

With this back-end architecture we ensure reliability, since every necessary data is stored in the database, from session information to the received sensor data, before sending the response to the gathering unit.

Our server is designed to store the data in a local PostgreSQL database, which is configured to replicate the data to a secondary server.
We considered using noSQL since it has some advantages over relational databases on huge datasets, mainly with single-index write operations. 
However, they have a similar performance in multiple-index multiple-writes, and even have lower performance on complex reads \cite{6253535}.
We opted for a relational database since our system performs mainly multiple-index bulk writes to the servers and can have a huge amount of multiple-index complex reads.

Our data processing projects and front-end visualizations typically query the database using multiple indexes, such as by session\_id and timestamp when showing a session to its user, by GPS coordinates for spatial queries or clustering, and many others.
These queries may also require correlating the data between many sensors, such as when mapping Wifi hotspots as can be seen in our examples in Section~\ref{sec:usecase}, making relational databases a perfect choice.
We opted for PostgreSQL by default since it offers a robust GIS extension, and a large part of our data analysis is based on geographical information.
Furthermore, it allows an easy setup of replication or multi-master architectures, solving scalability problems that may occur.

\begin{figure}[tb]
 \centering
 \includegraphics[trim=1cm 6cm 1cm 1cm, clip=true,width=1\linewidth]{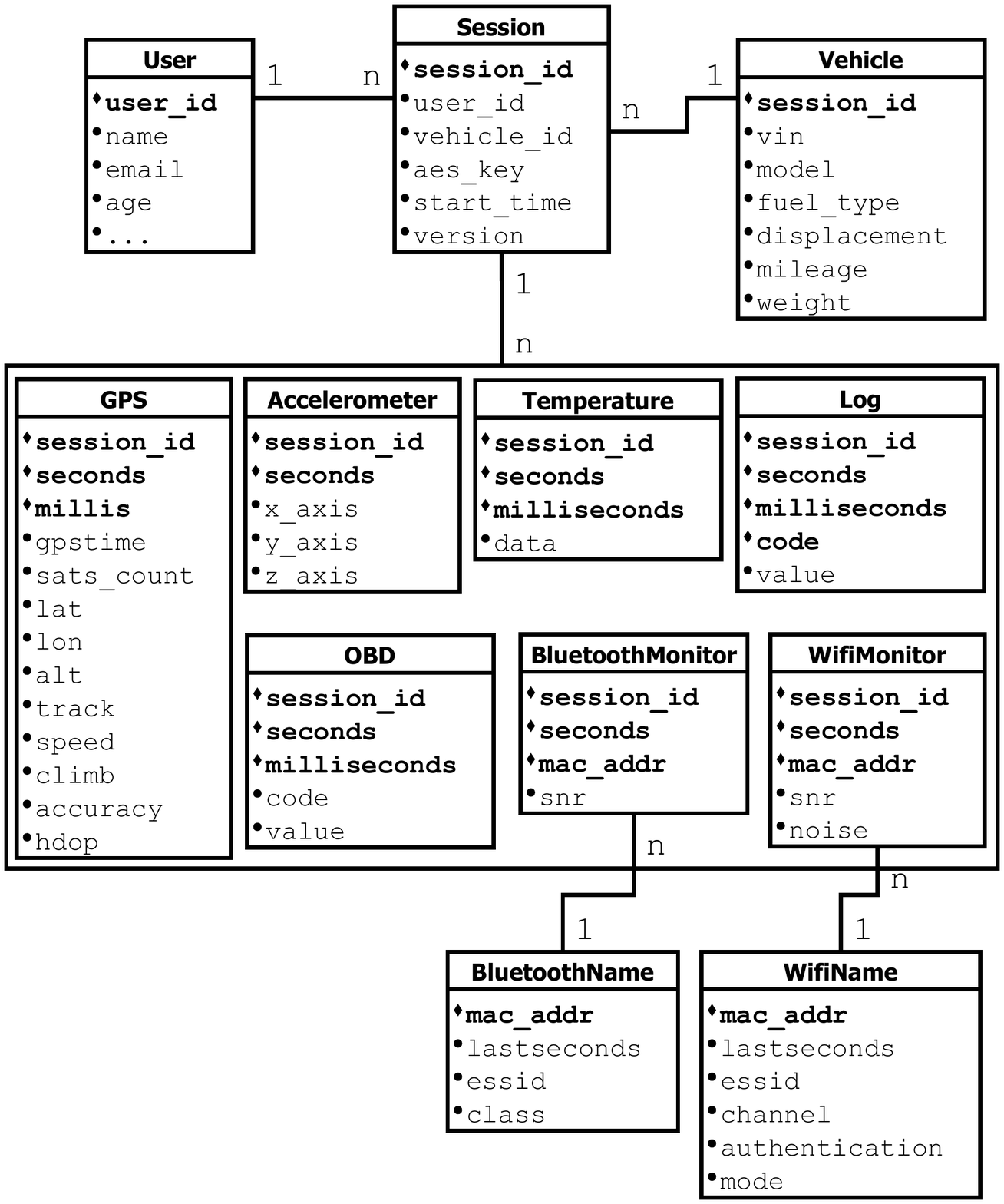}
 \caption{Database structure}
 \label{figure:Database}
\end{figure}


The implemented database structure can be seen in Fig.~\ref{figure:Database}, and follows the structure proposed in \ref{subsec:datastructure} with three distinct data types: Session information, gathered data, and auxiliary data.

The first stores Session information including vehicles' and users'.
As previously stated, for privacy reasons, the users' table is only accessible by the authentication mechanisms and by the database administrator.

The auxiliary data prevents the system from transmitting or storing redundant data, saving a lot of bandwidth and storage space.

We try to have a very similar structure for every table holding gathered data, with similar primary keys, codifications and column names.
For example, the tables storing accelerometer, gyroscope and magnetometer have exactly the same structure.

In Section~\ref{storagecosts} we present the storage requirements for each sensor used in common sensing tasks.

\subsubsection{Front End}
\label{subsec:frontend}

The gathered data can only be accessed and processed through the secondary servers.
We implemented a simple visualization web-page, that handles authentication via OpenID with a Google account, and allows a user to visualize all of his recorded sessions and associated data using a GIS service.
Other running services and algorithms are constantly providing new data to the secondary databases, such as calculating fuel consumption from \ac{OBD} data, or mapping the wifi signal strength of detected open networks.
Some of these projects that provide extra information are described in Section~\ref{sec:usecase}.

Besides the user visualization webpage, we provide a webpage that allows a user to access, download and delete all of his data.

\section{Meeting the User Interests}
\label{sec:meetinginterests}

\subsection{Saving the user's battery}

The reconfigurability of our system was leveraged to perform a battery consumption study with different classes of devices, sensor configurations and sampling rates.
In this section we compare a 2010 small and cheap smartphone - Samsung I5500 running Android version 2.3.7 with a single-core 600\,MHz CPU and a 1200\,mAh battery -  to a 2012 powerful model - LG Nexus 4 E960 running Android 4.2.2 with a quad-core 1.5\,GHz CPU and a 2100\,mAh battery.
The study involved at least 5 runs of each configuration and smartphone model, with everything turned off except for our running application and desired sensors, and analyzing the battery level (\%) after a 3 hours data-gathering session.
We consider these results, shown as percentage of battery consumed and estimated time to battery depletion, to be more informative to the users.

\begin{table}[tb]\scriptsize
 \centering
 \caption{Average battery consumption and autonomy after 3\,h sensing periods}
\begin{tabular}{c|c|c|c|c|}
\cline{2-5}
\multicolumn{1}{c|}{}                                                                      & \multicolumn{2}{|c|}{Samsung I5500} & \multicolumn{2}{|c|}{Nexus 4 E960} \\ \cline{2-5} 
\multicolumn{1}{c|}{}                                                                      & 3h Cons.          & Auton.         & 3h Cons.         & Auton.        \\ \hline
\multicolumn{1}{|c|}{Idle}                                                                 & 5.6\%            & 53.6h            & 1.4\%           & 214.3h          \\ \hline
\multicolumn{1}{|c|}{Acc, Gyro, Mag}                                                       & 16.7\%           & 18.0h            & 4.3\%           & 69.8h           \\ \hline
\multicolumn{1}{|c|}{GPS}                                                                  & 25.0\%           & 12.0h            & 14.0\%          & 21.4h           \\ \hline
\multicolumn{1}{|c|}{Wifi (WF) Scan}                                                            & 19.8\%           & 15.2h            & 11.3\%          & 26.5h           \\ \hline
\multicolumn{1}{|c|}{BlueTooth (BT) Scan}                                                              & 11.0\%           & 27.3h            & 6.0\%           & 50.0h           \\ \hline
\multicolumn{1}{|c|}{External via BT}                                                      & 11.0\%           & 27.3h            & 9.7\%           & 30.9h           \\ \hline
\multicolumn{1}{|c|}{Ambient pressure}                                                          &                  &                  & 5.0\%           & 60.0h           \\ \hline
\multicolumn{1}{|c|}{\begin{tabular}[c]{@{}c@{}}Acc, Gyro, Mag, GPS\end{tabular}}         & 34.8\%           & 8.6h             & 16.6\%          & 18.1h           \\ \hline
\multicolumn{1}{|c|}{\begin{tabular}[c]{@{}c@{}}Acc, Gyro, Mag\\ GPS, WF, BT\end{tabular}} & 49.8\%           & 6.0h             & 24.3\%          & 12.4h           \\ \hline
\multicolumn{1}{|c|}{All internal}                                                         & 49.8\%           & 6.0h             & 28.4\%          & 10.6h           \\ \hline
\end{tabular}
\label{table:battery}
\end{table}

\begin{figure}[tb]
 \centering
 \includegraphics[width=1\linewidth]{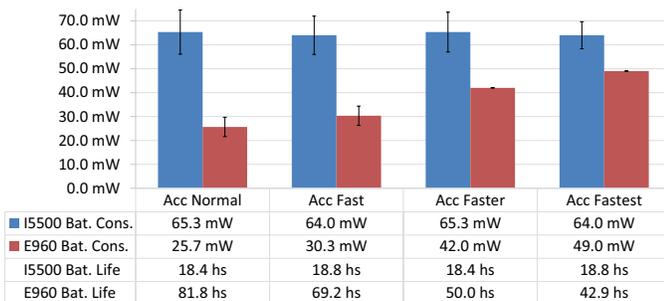}
 \caption{Energy consumption for different Accelerometer sampling rates}
 \label{figure:AccelBattery}
\end{figure}

Table~\ref{table:battery} presents some of these results, where we can see a clear improvement between phones on the energy consumption of the Accelerometer, Gyroscope and Magnetometer sensors, even after taking into consideration the differences in battery capacity (see estimated wattage from Fig.~\ref{figure:AccelBattery}).

\begin{figure}[tb]
 \centering
 \includegraphics[width=1\linewidth]{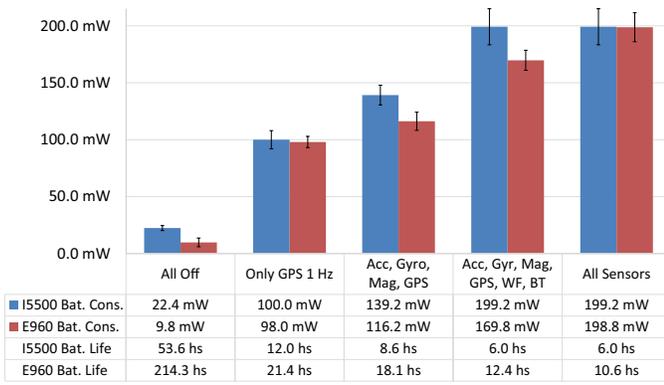}
 \caption{Energy consumption for different sensor setups}
 \label{figure:SetupsBattery}
\end{figure}

Fig.~\ref{figure:SetupsBattery} presents the most common sensing setups and their corresponding battery consumption in average watts per hour (Bat. Cons.) and estimated battery autonomy (Bat. Life).
By performing this analysis we can give the users and researchers a prediction of the battery requirements of a specific sensing scenario.

Fig.~\ref{figure:AccelBattery} compares the battery consumption of different accelerometer sampling rates.
The older Samsung phone showed a constant consumption across all sampling rates, from 5\,Hz to 50\,Hz, while the newer Nexus 4 showed a more efficient sensor but with an increasing consumption for higher sampling rates, from 5\,Hz to 200\,Hz.

\begin{figure}[tb]
 \centering
 \includegraphics[width=1\linewidth]{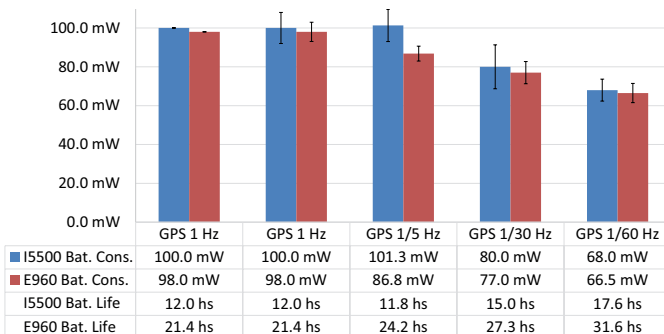}
 \caption{Energy consumption for different GPS sampling rates}
 \label{figure:GPSBattery}
\end{figure}

From Fig.~\ref{figure:GPSBattery}, we can see that the GPS energy consumption does not scale linearly with the sampling rate, which is expected since a GPS fix takes at least a few seconds to obtain after waking up from standby.
A 60\,s GPS sampling rate results in a 35\% of energy reduction (-35\,mWh) comparing to a 1\,s sampling.

\begin{figure}[tb]
 \centering
 \includegraphics[width=1\linewidth]{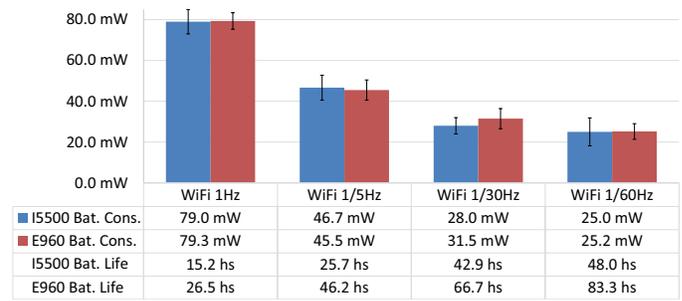}
 \caption{Energy consumption for different WiFi sampling rates}
 \label{figure:WiFiBattery}
\end{figure}

In a similar fashion, we can see from Fig.~\ref{figure:WiFiBattery} that just like the GPS, the WiFi module energy consumption does not scale proportionally to the sampling rate.
However, we can save more energy from this sensor than the GPS, saving 69\% (-54\,mWh) by using a 60\,s sampling rate instead of 1\,s.

Overall, we can see that, in the tested models, the smartphones' battery lasts for 6\,h to 10\,h when collecting data from every internal sensor.
A user that travels for 1\,h every day will arrive at the end of the day with 15\% less battery for sensing those trips.

\subsection{Minimizing the user's storage and communication costs}
\label{storagecosts}

The size and amount of collected data can have a big impact on the cost-of-knowledge, depending on the communication and storage costs.
To save storage space, our mobile application compresses the gathered data and stores higher-frequency sensors in arrays.
In Table~\ref{table:storage} we present the resulting storage requirement of individual sensors and typical configurations.

\begin{table}[tb]\scriptsize
 \caption{Storage requirements for different sensors}
 \centering
 \begin{tabular}{|c|c|c|c|}
 \hline
Streams &	Sample size &	Sample rate &	Approx storage\\
 \hline
GPS &				47\,B &	1\,Hz &			50\,B/s\\
 \hline
Acc, Gyro, Mag &			8+3x2\,B &		4\,-\,200\,Hz &		32\,-\,1200,B/s\\
 \hline
Wifi Scanner &		17\,B &		0-10\,Hz &		10\,-\,100\,B/s\\
 \hline
BT Scanner &		16\,B &		0-20\,/10\,s &	0\,-\,20\,B/s\\
 \hline
Pressure &			12\,B &		1\,/10\,s &		1\,B/s\\
 \hline
OBD &				14\,B &		4-12\,Hz &		112\,B/s\\
 \hline
 \hline
 \multicolumn{2}{|l|}{Typical storage} & \multicolumn{2}{|c|}{$\simeq$\,150\,Bps}\\
 \hline
 \multicolumn{2}{|l|}{Storage with all sensors} & \multicolumn{2}{|c|}{\textless\,400\,Bps}\\
 \hline
 \multicolumn{2}{|l|}{Maximum storage} & \multicolumn{2}{|c|}{$\simeq$\,3800\,Bps}\\
 \hline

 \end{tabular}
 \label{table:storage}
\end{table}

With default configurations, our application is gathering data from GPS at 1Hz; Wifi scanner sensing networks every 2 or 3 seconds; and Accelerometer, Gyroscope and Magnetometer at the default Android rate, typically between 4 and 15\,Hz.
This configuration results in around 150\,bytes of data stored per second, or 540\,KB per hour, already including identifiers and timestamps.

In a more sensing intensive utilization, gathering data with the highest available sample rate, with many surrounding wifi networks and gathering vehicular data from an external OBD device, the bandwidth requirement is around 4\,KB of data per second, 13\,MB per hour.
The accelerometer, gyroscope and magnetometer are responsible for more than 90\% of the storage utilization.

Our mobile application uses SQLite to store data before being transmitted, which requires some overhead storage.
Tests showed a storage efficiency between 70\% and 95\% depending on the size and sensors configuration.
Also, before transmission, our protocol serializes the data in JSON format, compresses using zlib, encrypts and transmits the data to the server.
Our tests indicate that a 1\,h session occupying 700\,KB in sqlite storage, are encapsulated in a JSON string with around 1100\,KB of length, but after our protocol's compression and encryption result in just 120\,KB of transmitted data (83\% compression from the SQLite storage).


\subsection{Meeting the privacy expectations of the user}

Studies indicate it is very hard to provide true anonymity and privacy when storing users' location history~\cite{Emiliano}, but there are some ways to improve it.

To this end, our system implements the following techniques:
\begin{itemize}
\item Every communication is secured to prevent eavesdroppers from obtaining possibly sensitive information;
\item We use third-party openID authentication, preventing us from handling and storing users' passwords;
\item Only the user's hashed email address is stored in the DB for authentication purposes, but it is not accessible by any service nor researcher;
\item Users retain full ownership of gathered data, and our frontend allows a user to completely delete his sessions if desired.
\end{itemize}

We have analyzed the European Union Data Protection Guidelines, and our proposed architecture conforms with the suggested privacy and anonymity features, and provides mechanisms that allow services and applications built upon it to do as well.
\section{Experimental Results}
\label{sec:usecase}

We have been running our system for a few months, and so far we have gathered around 150 million rows of data from 115 different volunteers, 3.500 trips and a total of more than 6.200 hours of sensing time.
From those we have 8 million GPS points, 22 million seconds of accelerometer data at approximately 5Hz, 3 millions of OBD vehicle data points, and 57 million wifi signal strength points from 295.000 different access points, 69.900 of which are open without protection.

\begin{figure}[tb]
 \centering
 \includegraphics[width=1.0\linewidth]{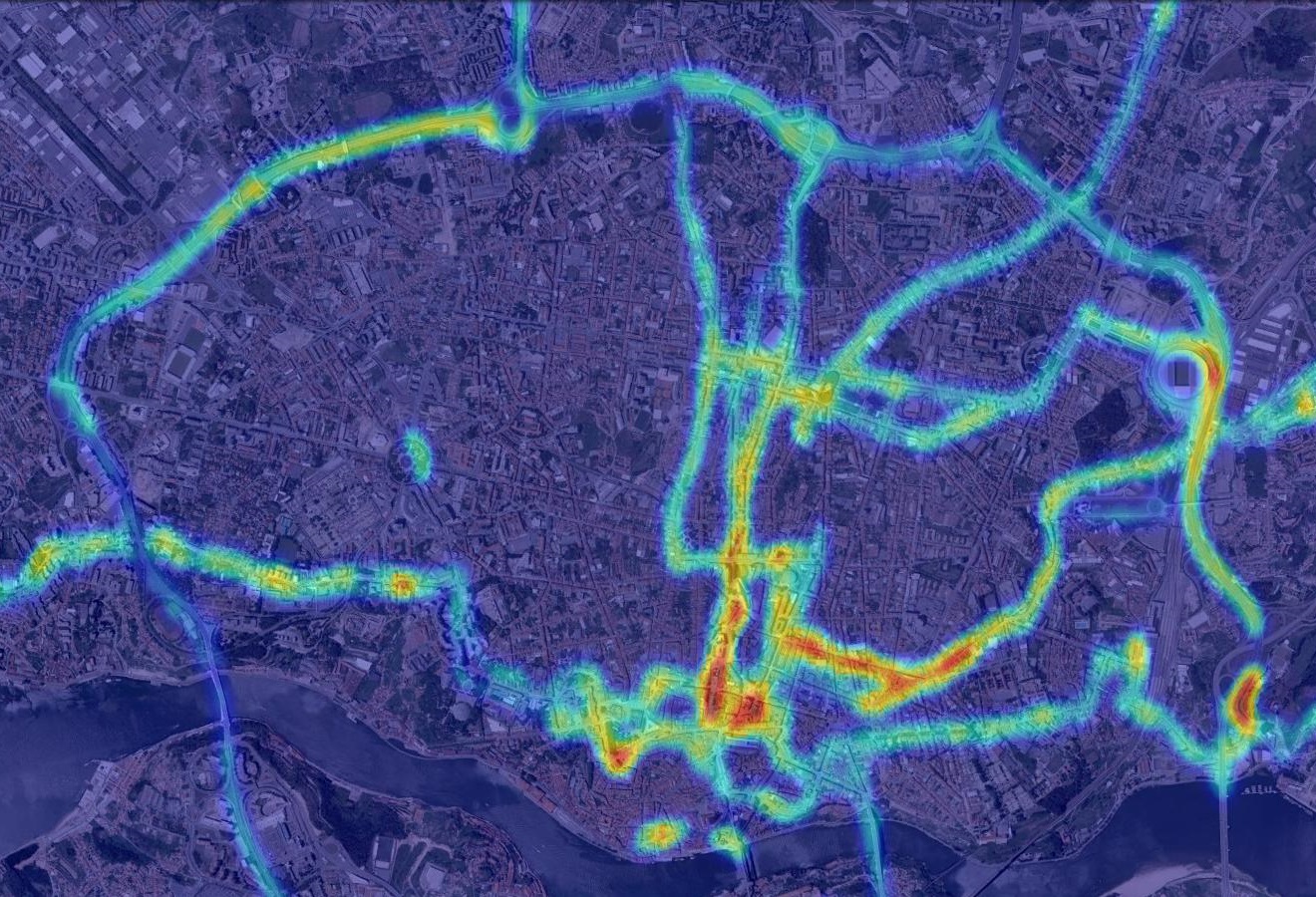}
 \caption{Cardiac stress levels in Porto's BUS Drivers}
 \label{figure:portostress}
\end{figure}

One of the projects leveraging our platform aimed at cardiac stress detection among public bus drivers~\cite{Rodrigues2014}.
The bio-monitoring external devices supported by our app can gather cardiac information, which was used to estimate stress and create the stress map shown in Fig.~\ref{figure:portostress}.


\begin{figure}[tb]
 \centering
 \includegraphics[width=1.0\linewidth]{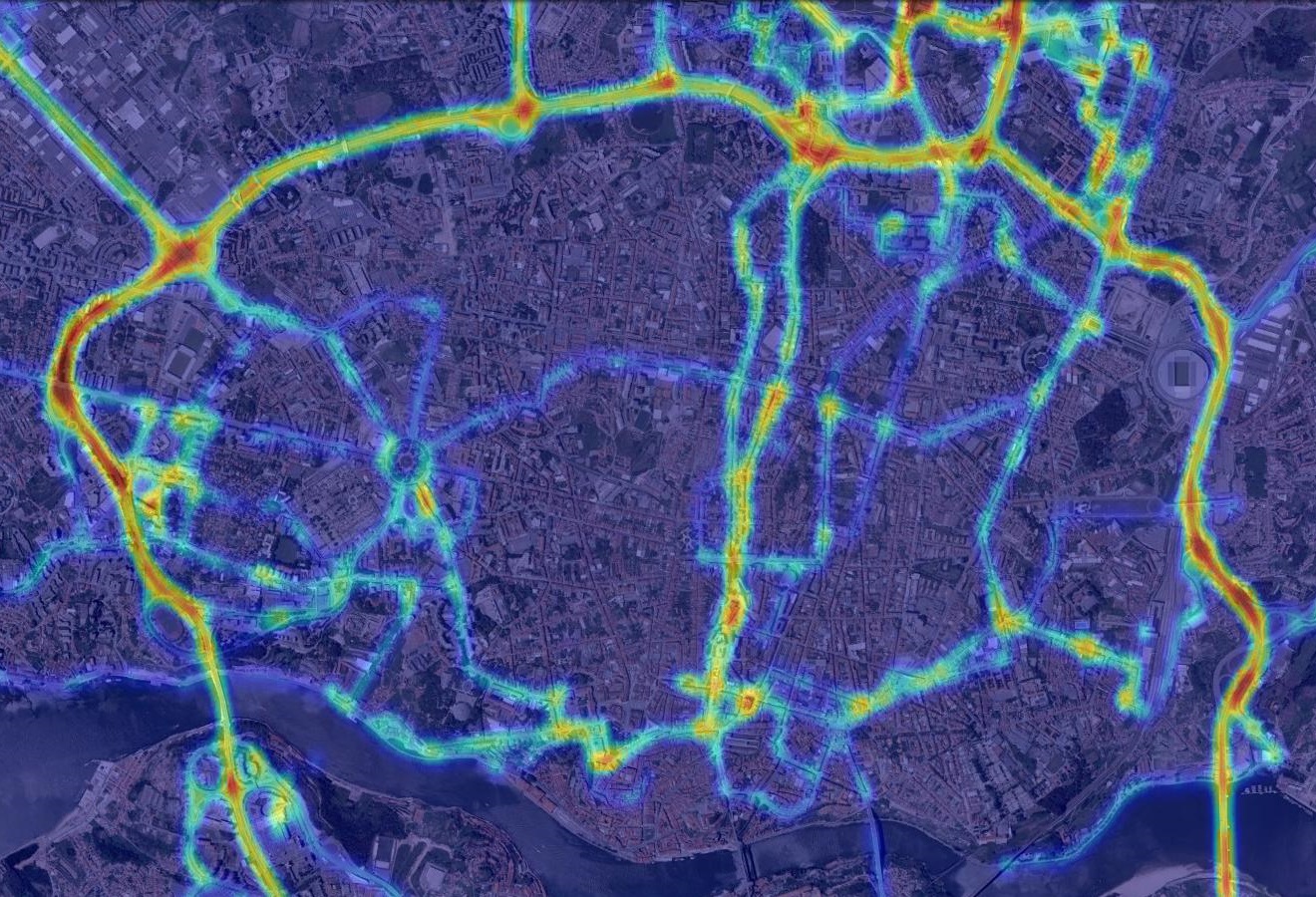}
 \caption{Fuel Consumption among Porto's SenseMyCity Users}
 \label{figure:portofc}
\end{figure}

Our platform was also used to connect to \ac{OBD} external sensors and gather vehicle information such as speed, fuel consumption and other driving metrics.
The map in Fig.~\ref{figure:portofc} was created by clustering fuel consumption data in the city of Porto.
The data gathered from the \ac{OBD} external sensors was used to improve fuel consumption estimation in smartphones without requiring any external sensor~\cite{Ribeiro2013}.
We used the synchronized data from the OBD and GPS, to train a mathematical model to estimate the instantaneous fuel consumption based solely on GPS data: vehicle speed, acceleration, and road gradient.



Another use case aims at improving user’s or vehicle’s WiFi connectivity in urban scenarios.
A stable internet access can greatly improve a user’s experience and allow for real time synchronization of data.
The project analyzes the georeferenced Wifi data collected to estimate the location of WiFi access points (AP).
Furthermore, connectivity is optimised by predicting the optimum sequence of AP for a given travel path that minimizes the amount of handshakes and maximize availability and path coverage.



SenseMyCity is also being used in some universities as a data provider for machine learning courses and projects.
Teachers and students leverage our system to gather a real world data-set tightly related to the user (student), and export their gathered data to a standard format.


\section{Related Work}
\label{sec:related}

Lane~\cite{Lane2010a} presents a very thorough survey of existing mobile phone sensing applications, strategies and policies.
Similarly, Chatzimilioudis~\cite{Chatzimilioudis} compares recent smartphone-based crowdsourcing applications and classifies them according to several aspects. 
The required user involvement is considered one of the most important aspects, which can be either participatory or opportunistic.
According to this classification, participatory systems require a user to report or subscribe to sensing tasks, aiming to gather data from targeted events or places in time and space.
On the other hand, opportunistic systems aim to gather a large number of raw data with almost no user interaction.
This model typically requires more offline data processing, to filter and infer information from the data.

Some systems have been developed that allow a quick integration of crowdsourcing technologies, such as the Ushahidi project~\cite{Okolloh}, designed to easily deploy a crowdsource mapping platform to collect reports from volunteers during natural disasters or other events.
These and similar participatory platforms, requiring users to actively participate in the sensing activity, have been used by a multitude of projects~\cite{Powell2012}\cite{ Dugdale}\cite{mcclendon2012leveraging}.
Participatory platforms can be used to quickly detect and report on main events, but fail to detect secondary events or unnoticeable patterns in the environment that the users do not consider worth reporting.
The projects Medusa~\cite{Medusa}, mCrowd~\cite{YanmCrowd} and the authors in~\cite{lasnia2010crowdsourcing} also present participatory crowdsensing platforms with user-in-the-loop designs.
Besides just gathering user reports, these also support the collection of raw sensor data as specified in the task, defined by other users or researchers as a sensing requirements, and the system distributes these tasks to smartphone users who wish to participate, while handling optional monetary incentives.

Incentives are an important aspect of such participatory systems as they require a large user interactivity.
They can be given in the form of monetary compensation~\cite{Medusa}, by providing services~\cite{Boulos2011}, by entertaining the user~\cite{Urbanopoly}, or even by motivating altruism~\cite{Chatzimilioudis}.

In an opportunistic way, StreetSmart~\cite{StreetSmart} uses crowdsensing with mobile phones to gather data from GPS, Accelerometer and OBD data, and provides fuel consumption information to users.
CarTel~\cite{Hull2006} has a similar approach, gathering information solely from Global Positioning System (GPS) and On-Board Diagnostics (OBD).
These projects implement sensing applications focused on solving problems for their specific scenarios, such as implementing delay-tolerant network communications, and gathering data from only a small subset of sensors.
Our platform was designed to be able to cope with a larger set of scenarios and requirements, by being easily configured and able to gather data from a multitude of sensors.
Fuel consumption estimation requires only a small subset of sensors and capabilities available in our platform. A similar calculation was inclusively performed over our gathered data~\cite{Ribeiro2013}.

With a completely different goal, the CrowdSense@Place framework~\cite{charplaces} uses opportunistically captured images and audio clips from smartphones to automatically identify and characterize places a user visits, without requiring any user interaction.

In this work we propose a more general and configurable opportunistic data gathering platform, that is able to gather data from a large number of sensors available in smartphones without hindering its normal operation or requiring regular conscious interactions from the user.
A lot of information can be retrieved from such massive data gathering system, such as estimating fuel consumption~\cite{Hull2006}\cite{Ribeiro2013} or monitoring health-care systems~\cite{Boulos2011}\cite{Rodrigues2014}.
Moreover, by aggregating collected data it is possible to infer much more information and provide useful services, such as automatic car sharing suggestions from traces similarity as proposed in~\cite{Zeinalipour}, predicting congestions and movement patterns~\cite{calabrese_ieeetits2011}\cite{Zimmerman2011}, estimating per street average fuel consumption and cost~\cite{StreetSmart} or predicting network connectivity from upcoming open WiFi access points (Section~\ref{sec:usecase}).

However, it is very hard to gather a large data-set without compromising the privacy and anonymity of the users.
Studies show that although people are generally very permissive regarding location privacy policies, it is largely dependent on how their location data is used and with whom it is shared~\cite{KrummSurvey}\cite{Tsai2009}.
There have been many proposals to integrate privacy in participatory systems~\cite{Emiliano}\cite{Hoh2010}\cite{Shilton2009}, mainly in systems offering services that allows a user to access and analyze data reported by others.

Crowdsensing platforms are vulnerable to misbehaving nodes and the existence of noisy or biased data, which can impact the quality of the information inferred from the gathered data~\cite{boulos2011crowdsourcing}.
In our platform, data reliability and misbehaving nodes should be addressed during the data processing phase~\cite{Tanachaiwiwat2005}.

\section{Conclusions}
\label{sec:conclusion}

Instead of implementing a platform for a specific scenario with specific limiting requirements, we designed a superseding and modular opportunistic system that can be used in distinct sensing projects and easily extended for new sensors.
The application is fully configurable and able to gather data while running on the background, while requiring minimal user interaction.
The many configuration parameters can, among others, choose the gathered sensors and their sampling rates, and make it transmit data in real-time while choosing allowed connection types (e.g. send only through WiFi).
It then uses backend servers to store and process the massive amounts of data according to the researchers and projects goals.

The reconfigurability and ubiquity nature of our mobile platform allows us to benchmark resources utilization in different smartphone models and configuration, providing us an estimation of battery consumption and storage requirements for some desired sensing task.
This way, besides providing valuable data-sets our platform allows researchers to estimate and minimize the cost-of-knowledge for different projects.

\section*{Acknowledgments}

This work was financed by Instituto de Telecomunica\c{c}\~{o}es,
by the European FP7 Project - Future Cities: FP7-REGPOT-2012-2013-1,
by National Funds through the FCT – Fundação para a Ciência e a Tecnologia 
within project PTDC/EEI-ELC/2760/2012 and grant SFRH/BD/62537/2009.

\bibliographystyle{IEEEtran}
\bibliography{MDD,biblio}

\begin{thebibliography}{10}
\providecommand{\url}[1]{#1}
\csname url@samestyle\endcsname
\providecommand{\newblock}{\relax}
\providecommand{\bibinfo}[2]{#2}
\providecommand{\BIBentrySTDinterwordspacing}{\spaceskip=0pt\relax}
\providecommand{\BIBentryALTinterwordstretchfactor}{4}
\providecommand{\BIBentryALTinterwordspacing}{\spaceskip=\fontdimen2\font plus
\BIBentryALTinterwordstretchfactor\fontdimen3\font minus
  \fontdimen4\font\relax}
\providecommand{\BIBforeignlanguage}[2]{{%
\expandafter\ifx\csname l@#1\endcsname\relax
\typeout{** WARNING: IEEEtran.bst: No hyphenation pattern has been}%
\typeout{** loaded for the language `#1'. Using the pattern for}%
\typeout{** the default language instead.}%
\else
\language=\csname l@#1\endcsname
\fi
#2}}
\providecommand{\BIBdecl}{\relax}
\BIBdecl

\bibitem{Boulos2011}
M.~N.~K. Boulos, S.~Wheeler, C.~Tavares, and R.~Jones, ``{How smartphones are
  changing the face of mobile and participatory healthcare: an overview, with
  example from eCAALYX.}'' \emph{Biomed. Eng. Online}, vol.~10, no.~1, p.~24,
  Jan. 2011.

\bibitem{Shilton2009}
K.~Shilton, J.~a. Burke, D.~Estrin, and M.~Hansen, ``{Designing the Personal
  Data Stream : Enabling Participatory Privacy in Mobile Personal Sensing},''
  \emph{Work}, no. September, pp. 25--27, 2009.

\bibitem{Murty2008a}
R.~N. Murty, G.~Mainland, I.~Rose, A.~R. Chowdhury, A.~Gosain, J.~Bers, and
  M.~Welsh, ``{CitySense: An Urban-Scale Wireless Sensor Network and
  Testbed},'' in \emph{2008 IEEE Conf. Technol. Homel. Secur.}\hskip 1em plus
  0.5em minus 0.4em\relax IEEE, May 2008, pp. 583--588.

\bibitem{Rodrigues2011}
J.~G.~P. Rodrigues, A.~Aguiar, F.~Vieira, J.~Barros, and J.~P.~S. Cunha, ``{A
  mobile sensing architecture for massive urban scanning},'' in \emph{2011 14th
  Int. IEEE Conf. Intell. Transp. Syst.}\hskip 1em plus 0.5em minus 0.4em\relax
  Ieee, Oct. 2011, pp. 1132--1137.

\bibitem{6253535}
J.~S. van~der Veen, B.~van~der Waaij, and R.~J. Meijer, ``{Sensor Data Storage
  Performance: SQL or NoSQL, Physical or Virtual},'' in \emph{Cloud Comput.
  (CLOUD), 2012 IEEE 5th Int. Conf.}, 2012, pp. 431--438.

\bibitem{Emiliano}
E.~D. Cristofaro and C.~Soriente, ``{Participatory Privacy: Enabling Privacy in
  Participatory Sensing},'' \emph{CoRR}, vol. abs/1201.4, 2012.

\bibitem{Rodrigues2014}
J.~G.~P. Rodrigues, M.~Kaiseler, A.~Aguiar, J.~Barros, and J.~P.~S. Cunha, ``{A
  Mobile Sensing Approach to Stress Detection and Memory Activation for Public
  Bus Drivers},'' \emph{Submitt. Publ.}, 2014.

\bibitem{Ribeiro2013}
V.~Ribeiro, J.~Rodrigues, and A.~Aguiar, ``{Mining geographic data for fuel
  consumption estimation},'' in \emph{16th Int. IEEE Conf. Intell. Transp.
  Syst. (ITSC 2013)}.\hskip 1em plus 0.5em minus 0.4em\relax IEEE, Oct. 2013,
  pp. 124--129.

\bibitem{Lane2010a}
N.~D. Lane, E.~Miluzzo, H.~Lu, D.~Peebles, T.~Choudhury, A.~T. Campbell, and
  CollegeDartmouth, ``{Adhoc And Sensor Networks: A Survey of Mobile Phone
  Sensing},'' \emph{IEEE Commun. Mag.}, no. September, pp. 140--150, 2010.

\bibitem{Chatzimilioudis}
G.~Chatzimilioudis, A.~Konstantinidis, C.~Laoudias, and D.~Zeinalipour-Yazti,
  ``{Crowdsourcing with Smartphones},'' \emph{Internet Comput. IEEE}, vol.~16,
  no.~5, pp. 36--44, 2012.

\bibitem{Okolloh}
O.~Okolloh, ``{Ushahidi, or 'testimony': Web 2.0 tools for crowdsourcing crisis
  information},'' \emph{Particip. Learn. Action}, vol.~59, no.~1, pp. 65--70,
  2009.

\bibitem{Powell2012}
J.~Powell, G.~Nash, and P.~Bell, ``{GeoExposures: Documenting temporary
  geological exposures in Great Britain through a citizen-science web site},''
  \emph{Proc. Geol. Assoc.}, no.~0, pp.~--, 2012.

\bibitem{Dugdale}
J.~Dugdale, B.~de~Walle, and C.~Koeppinghoff, ``{Social media and SMS in the
  haiti earthquake},'' in \emph{Proc. 21st Int. Conf. companion World Wide
  Web}, ser. WWW '12 Companion.\hskip 1em plus 0.5em minus 0.4em\relax New
  York, NY, USA: ACM, 2012, pp. 713--714.

\bibitem{mcclendon2012leveraging}
S.~McClendon and A.~C. Robinson, ``{Leveraging Geospatially-Oriented Social
  Media Communications in Disaster Response},'' 2012.

\bibitem{Medusa}
M.-R. Ra, B.~Liu, T.~F. {La Porta}, and R.~Govindan, ``{Medusa: a programming
  framework for crowd-sensing applications},'' in \emph{Proc. 10th Int. Conf.
  Mob. Syst. Appl. Serv.}, ser. MobiSys '12.\hskip 1em plus 0.5em minus
  0.4em\relax New York, NY, USA: ACM, 2012, pp. 337--350.

\bibitem{YanmCrowd}
T.~Yan, M.~Marzilli, R.~Holmes, D.~Ganesan, and M.~Corner, ``{mCrowd: a
  platform for mobile crowdsourcing},'' in \emph{Proc. 7th ACM Conf. Embed.
  Networked Sens. Syst.}, ser. SenSys '09.\hskip 1em plus 0.5em minus
  0.4em\relax New York, NY, USA: ACM, 2009, pp. 347--348.

\bibitem{lasnia2010crowdsourcing}
D.~Lasnia, A.~Broering, S.~Jirka, and A.~Remke, ``{Crowdsourcing Sensor Tasks
  to a Socio-Geographic Network},'' in \emph{13th Agil. Int. Conf. Geogr. Inf.
  Sci.}, 2010.

\bibitem{Urbanopoly}
I.~Celino, D.~Cerizza, S.~Contessa, M.~Corubolo, D.~Dell'Aglio, E.~D. Valle,
  and S.~Fumeo, ``{Urbanopoly -- A Social and Location-Based Game with a
  Purpose to Crowdsource Your Urban Data},'' in \emph{Privacy, Secur. Risk
  Trust (PASSAT), 2012 Int. Conf. 2012 Int. Confernece Soc. Comput.}, 2012, pp.
  910--913.

\bibitem{StreetSmart}
A.~L. Oehlerking, ``{StreetSmart : modeling vehicle fuel consumption with
  mobile phone sensor data through a participatory sensing framework},''
  Master's thesis, Massachusetts Institute of Technology. Dept. of Mechanical
  Engineering., 2011.

\bibitem{Hull2006}
B.~Hull, V.~Bychkovsky, Y.~Zhang, K.~Chen, M.~Goraczko, A.~Miu, E.~Shih,
  H.~Balakrishnan, and S.~Madden, ``{CarTel: a distributed mobile sensor
  computing system},'' \emph{Proc. 4th Int. Conf. Embed. networked Sens.
  Syst.}, pp. 125--138, Oct. 2006.

\bibitem{charplaces}
Y.~Chon, N.~D. Lane, F.~Li, H.~Cha, and F.~Zhao, ``{Automatically
  characterizing places with opportunistic crowdsensing using smartphones},''
  in \emph{Proc. 2012 ACM Conf. Ubiquitous Comput.}, ser. UbiComp '12.\hskip
  1em plus 0.5em minus 0.4em\relax New York, NY, USA: ACM, 2012, pp. 481--490.

\bibitem{Zeinalipour}
D.~Zeinalipour-Yazti, C.~Laoudias, C.~Costa, M.~Vlachos, M.~Andreou, and
  D.~Gunopulos, ``{Crowdsourced Trace Similarity with Smartphones},'' p.~1.

\bibitem{calabrese_ieeetits2011}
F.~{}Calabrese, M.~{}Colonna, P.~{}Lovisolo, D.~{}Parata, and C.~{}Ratti,
  ``{Real-Time Urban Monitoring Using Cell Phones: A Case Study in Rome},''
  \emph{IEEE Trans. Intell. Transp. Syst.}, vol.~12, no.~1, pp. 141--151, 2011.

\bibitem{Zimmerman2011}
J.~Zimmerman, A.~Tomasic, C.~Garrod, D.~Yoo, C.~Hiruncharoenvate, R.~Aziz,
  N.~R. Thiruvengadam, Y.~Huang, and A.~Steinfeld, ``{Field trial of
  Tiramisu},'' in \emph{Proc. 2011 Annu. Conf. Hum. factors Comput. Syst. - CHI
  '11}.\hskip 1em plus 0.5em minus 0.4em\relax New York, New York, USA: ACM
  Press, May 2011, p. 1677.

\bibitem{KrummSurvey}
J.~Krumm, ``{A survey of computational location privacy},'' \emph{Pers.
  Ubiquitous Comput.}, vol.~13, no.~6, pp. 391--399, 2009.

\bibitem{Tsai2009}
J.~Y. Tsai, P.~Kelley, P.~Drielsma, L.~F. Cranor, J.~Hong, and N.~Sadeh,
  ``{Who's viewed you?: The impact of feedback in a mobile location-sharing
  application},'' \emph{Proc. 27th Int. Conf. Hum. factors Comput. Syst. - CHI
  '09}, p. 2003, 2009.

\bibitem{Hoh2010}
B.~H.~B. Hoh, M.~Gruteser, H.~X.~H. Xiong, and a.~Alrabady, ``{Achieving
  Guaranteed Anonymity in GPS Traces via Uncertainty-Aware Path Cloaking},''
  \emph{IEEE Trans. Mob. Comput.}, vol.~9, no.~8, pp. 1089--1107, 2010.

\bibitem{boulos2011crowdsourcing}
M.~N.~K. Boulos, B.~Resch, D.~N. Crowley, J.~G. Breslin, G.~Sohn, R.~Burtner,
  W.~A. Pike, E.~Jezierski, and K.-Y.~S. Chuang, ``{Crowdsourcing, citizen
  sensing and sensor web technologies for public and environmental health
  surveillance and crisis management: trends, OGC standards and application
  examples},'' \emph{Int. J. Health Geogr.}, vol.~10, no.~1, p.~67, 2011.

\bibitem{Tanachaiwiwat2005}
S.~Tanachaiwiwat and A.~Helmy, ``\BIBforeignlanguage{English}{{Correlation
  analysis for alleviating effects of inserted data in wireless sensor
  networks}},'' in \emph{\BIBforeignlanguage{English}{Second Annu. Int. Conf.
  Mob. Ubiquitous Syst. Netw. Serv.}}\hskip 1em plus 0.5em minus 0.4em\relax
  IEEE, 2005, pp. 97--108.

\end{thebibliography}

\begin{IEEEbiography}
[{\includegraphics[width=1in,height=1.25in,clip,keepaspectratio]{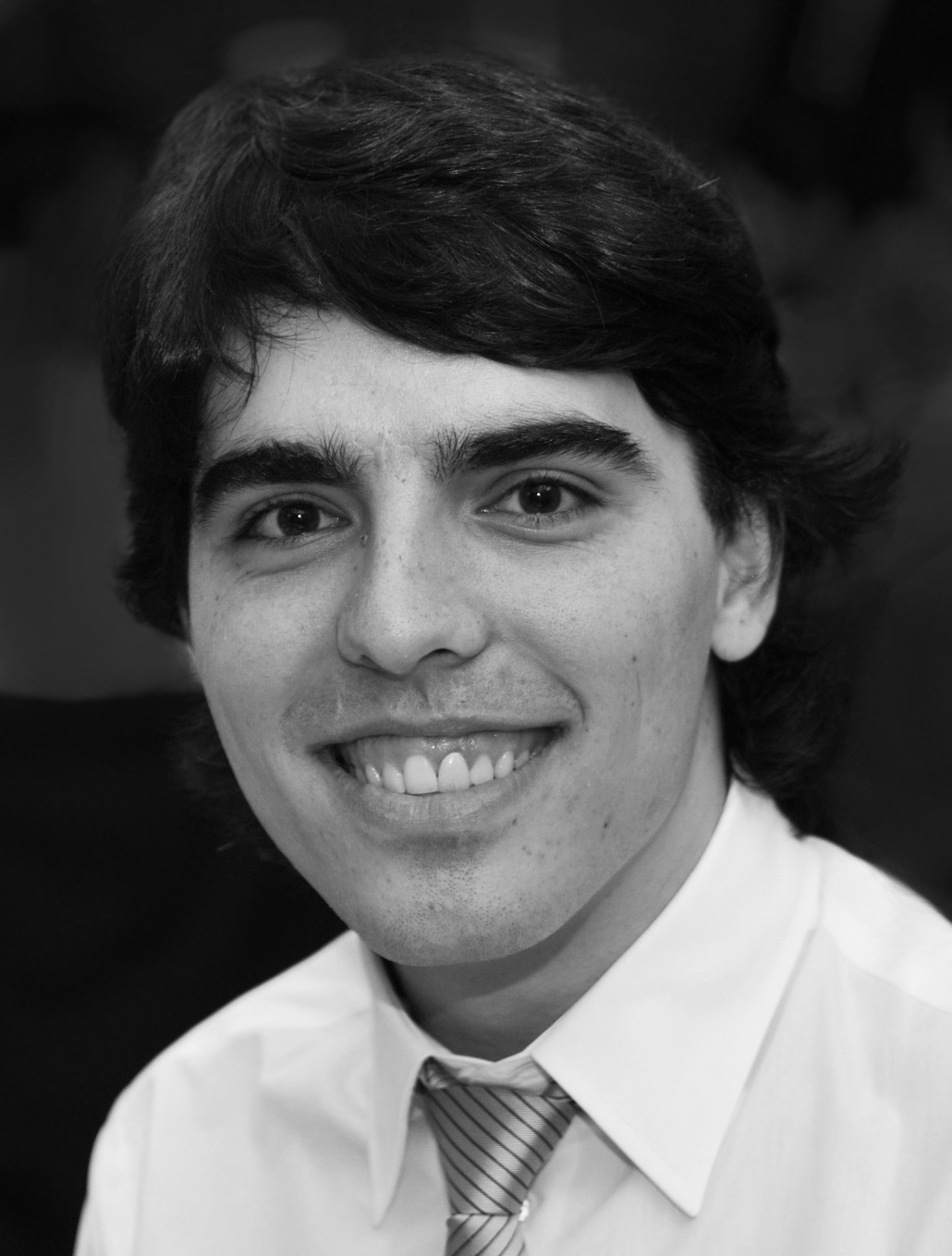}}]
{Jo\~ao Rodrigues} (S'11) received his M.Sc. degree in Electrical and Computer Engineering from the University of Porto, Portugal, in 2009, and since then is pursuing the Ph.D. degree at the same institution.
He develops his work at the Institute for Telecommunications (IT), and the main topics of his thesis are data gathering and mining in intelligent transportation systems.
He was awarded a Doctoral Scholarship from the Portuguese Foundation for Science and Technology in 2009.
His general interests are sensor networks and intelligent transportation systems.
\end{IEEEbiography}

\begin{IEEEbiography}
[{\includegraphics[width=1in,height=1.25in,clip,keepaspectratio]{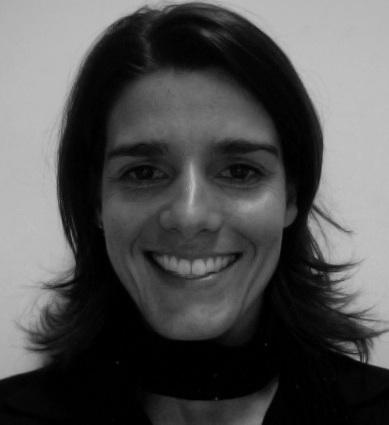}}]
{Ana Aguiar} (S'94-M'98-S'02-M'09) graduated in Electrical and Computer Engineering from the Faculty of Engineering University of Porto (FEUP), Portugal, in 1998, and received her PhD in Telecommunication Networks from the Technical University of Berlin, Germany, in 2008. She is an assistant professor at FEUP since 2009, with research interests in wireless networking and mobile sensing systems, specifically vehicular networks, crowd sensing, and machine-to-machine communications. She also contributes to several inter-disciplinary projects in the fields of intelligent transportation systems and well-being (stress). She began her career as an RF engineer working for cellular operators, and she worked at Fraunhofer Portugal AICOS on service-oriented architectures and wireless technologies applied to ambient assisted living. She has published in and is reviewer for several IEEE and ACM conferences and journals.
\end{IEEEbiography}

\begin{IEEEbiography}
[{\includegraphics[width=1in,height=1.25in,clip,keepaspectratio]{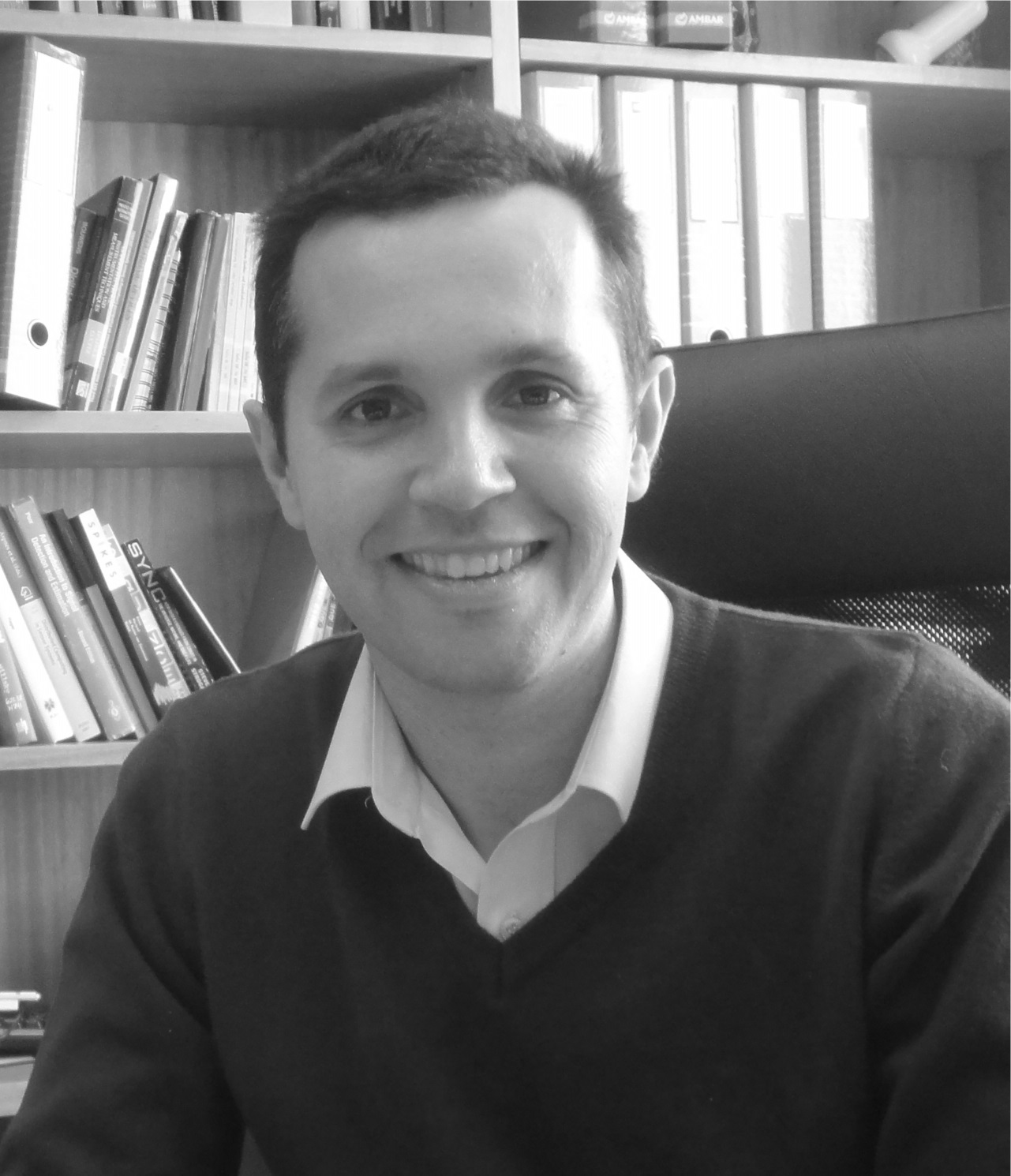}}]
{Jo\~ao Barros} (S'98-M'04-SM'11) is an Associate Professor of Electrical and Computer Engineering at the University of Porto and Founding Director of the Institute for Telecommunications (IT) in Porto, Portugal.
He also teaches at the Porto Business School and co-founded two recent startups, Streambolico and Veniam, commercializing wireless video and vehicular communication technologies, respectively.
He received his undergraduate education in Electrical and Computer Engineering from the Universidade do Porto (UP), Portugal and Universitaet Karlsruhe, Germany, and the Ph.D. degree in Electrical Engineering and Information Technology from the Technische Universitaet Muenchen (TUM), Germany.
\end{IEEEbiography}


\end{document}